\documentclass[jpcbfk,reprint,showkeys]{revtex4-1}

\renewcommand{\vec}[1]{\mbox{\boldmath$#1$}}

\newcommand{\dif}{\mathrm{d}}

\usepackage{graphicx}

\usepackage{amsmath}

\usepackage{amsfonts}

\usepackage{graphicx}

\begin{document}
\title{Molecular theory and the effects of solute attractive forces on 
hydrophobic interactions }
\author{Mangesh I. Chaudhari}
\email{michaud@sandia.gov}

\author{Susan B. Rempe}
\email{slrempe@sandia.gov}
\affiliation{Center for Biological and Material Sciences, Sandia National Laboratories,  
Albuquerque, NM 87185}

\author{D. Asthagiri}
\email{Dilip.Asthagiri@rice.edu}
\affiliation{Department of Chemical and Biomolecular Engineering, Rice
University, Houston, TX 77005}

\author{L. Tan}
\email{ltan2@tulane.edu}
\author{L. R. Pratt}
\email{lpratt@tulane.edu}
\affiliation{Department of Chemical and Biomolecular Engineering, Tulane
University, New Orleans, LA 70118}

\date{\today}

\begin{abstract} The role of solute attractive forces on hydrophobic
interactions is studied by coordinated development of theory and simulation
results for Ar atoms in water. We present a concise derivation of the local
molecular field (LMF) theory for the effects of solute attractive forces on
hydrophobic interactions, a derivation that clarifies the close relation of LMF
theory to the EXP approximation applied to this problem long ago. The simulation
results show that change from purely repulsive atomic solute interactions to
include realistic attractive interactions \emph{diminishes} the strength of
hydrophobic bonds. For the Ar-Ar rdfs considered pointwise, the numerical
results for the effects of solute attractive forces on hydrophobic interactions
are of opposite sign and larger in magnitude than predicted by LMF theory. That
comparison is discussed from the point of view of quasi-chemical theory, and it
is suggested that the first reason for this difference is the incomplete
evaluation within LMF theory of the hydration energy of the Ar pair. With a
recent suggestion for the system-size extrapolation of the required correlation
function integrals, the Ar-Ar rdfs permit evaluation of osmotic second virial
coefficients $B_2$. Those $B_2$ also show that incorporation of attractive
interactions leads to more positive (repulsive) values. With attractive
interactions in play, $B_2$ can change from positive to negative values with
increasing temperatures. This is consistent with the historical work of
Watanabe, \emph{et al.,} that $B_2 \approx 0$ for intermediate cases. In all
cases here, $B_2$ becomes more attractive with increasing temperature.
\end{abstract}

\maketitle

\section{Introduction}

The concept of a hydrophobic interaction is firmly embedded in general views of
the folding of water-soluble protein molecules. Kauzmann \cite{Kauzmann:1987uv}
clearly articulated that idea: ``Thus, proteins are stabilized by the same
physical forces as those that keep oil and water from mixing \ldots '' A key
phenomenological point is that enthalpic hydration contributions to the
thermodynamics of protein unfolding decrease, even vanish at moderate
temperatures \cite{Kauzmann:1987uv}. Hydrophobic interactions are then entropy
dominated. The enthalpy-entropy balance depends importantly on temperature, and
switches at higher temperatures \cite{Kauzmann:1987uv}. The entropic hydration
contributions to the thermodynamics of protein unfolding can vanish at higher
temperatures \cite{Kauzmann:1987uv}, and that condition has been called the
``entropy convergence'' point \cite{Garde:1996p7972}. Nevertheless, below such
an entropy convergence temperature, \emph{i.e.,} where hydrophobic
low-solubility is associated a negative entropy change, hydrophobic interactions
get stronger with temperature increases though with reduced rate of
strengthening.

From early days, the involvement of the hydration entropy has been
conceptualized by imagining \emph{icebergs} surrounding simple hydrophobic
solutes, such as inert gases, \emph{e.g.,} Ar below. Tanford \cite{Tanford:97}
attributed the original ``iceberg'' language to G. N. Lewis. Silverstein
\cite{Silverstein:2008va}, decades later, provides a modern view of the
relevance of the iceberg concept to hydrophobic phenomena. ``Iceberg'' is a
widely recognized figure of speech, but has not been the basis of defensible
statistical mechanical theory of these entropy effects. In fact, the statistical
mechanical theory that eventually does explain the entropy convergence phenomena
does not define or characterize iceberg structures \cite{Garde:1996p7972}. 

Because the iceberg parlance is vague and provocative, direct experimental
demonstrations of so-called inverse-temperature behaviors are particularly
helpful. Aggregation of sickle hemoglobin is a standard example
\cite{murayama1966molecular}. Well-known aqueous polymers that separate with
temperature \emph{increases,} \emph{i.e.,} systems that exhibit a lower critical
solution temperature (LCST), also provide examples. Elastin-like peptides (ELPs)
are probably the best known cases
\cite{Urry:1991ud,Urry:1982uy,Nuhn:2008jh,Reiersen:1998wb,Nicolini:2004dn}.
Substantial molecular simulation work is available describing ELP collapse
\cite{Li:2001ei,Li:2001cj,Li:2002dk,Rousseau:2004iu,Schreiner:2004ju,MarcelBaer:2006bu,Krukau:2007bc,Li:2014gq}
without addressing the statistical mechanical theory of hydrophobic
interactions. Those descriptive simulation efforts are largely consistent with
the traditional idea of the folding of elastin-like peptides upon heating, and
with each other, but not entirely \cite{Krukau:2007bc}.

Aqueous solutions of poly(N-isopropylacrylamide)s provide other examples of
LCSTs \cite{Maeda:2000ki,Qiu:2007kz}. Polyethylene glycols (PEGs) in water also
exhibit LCSTs \cite{Bae:1993uj}. The polymers noted are water-soluble below
their LCSTs. Thus they are substantially hydrophilic. But though they are
complicated molecules, they directly demonstrate the inverse-temperature
phenomena of classic hydrophobic effects.

The successful statistical mechanical theory for the entropy convergence
behavior
\cite{ghumm96,Gomez:1999hm,AshbaughHS:Colspt,hashb07,PrattLR:Molthe,pratt1999,Hummer:1998ij,PrattLR:Hydeam}
developed over decades from counter-intuitive initial steps
\cite{Pierotti:1967ux,Pierotti:1976tg,Pierotti:2001tj,BenNaim:1967ug,Stillinger:73,PRATTLR:Thehe,Chan:79,Rossky:1980vh,Pratt85}.
The statistical mechanical theory of hydrophobic interactions
\cite{PRATTLR:Thehe,chaudhari2013molecular} was formulated for hard sphere
hydrophobic solutes in water, and theoretical progress has been associated with
attention to detail for such simple cases
\cite{PRATTLR:Thehe,AshbaughHS:Colspt,chaudhari2013molecular}. That methodical
analysis strategy permits clarity in isolating the features that are the
ultimate interest. An important accomplishment of recent work
\cite{chaudhari2013molecular} was then to prove numerically that rigorously
defined hydrophobic interactions between atomic-sized hard sphere solutes in
water also exhibit inverse-temperature behavior. Independently, new results for
broader solute models arrived at consistent conclusions for those cases
\cite{koga2013osmotic}. Building from those important accomplishments, the
present work investigates the theory for adding attractive inter-atomic forces
to those primitive cases.

The counter-intuitive ingredients of the statistical mechanical theory
\cite{Chan:79} together with apparent disagreement with some experiments
\cite{Tucker:1979wk,Rossky:1980vh,Pratt:1985to,BernalP:VAPSOH} that do involve
attractive forces, lead promptly to questions about the consequences of solute
attractive forces associated with simple hydrophobic solutes
\cite{PrattLR:Effsaf}. That issue has been broadly discussed several times over
the intervening years
\cite{PrattLR:Effsaf,Pratt:1985to,WATANABEK:MOLSOT,Smith:1992p14802,%
SMITHDE:Freeea,Asthagiri:2008in} without achieving a definitive solution. That
situation can now change on the basis of the new results for hydrophobic
interactions noted above.

\begin{figure}
\includegraphics[width=3.0in]{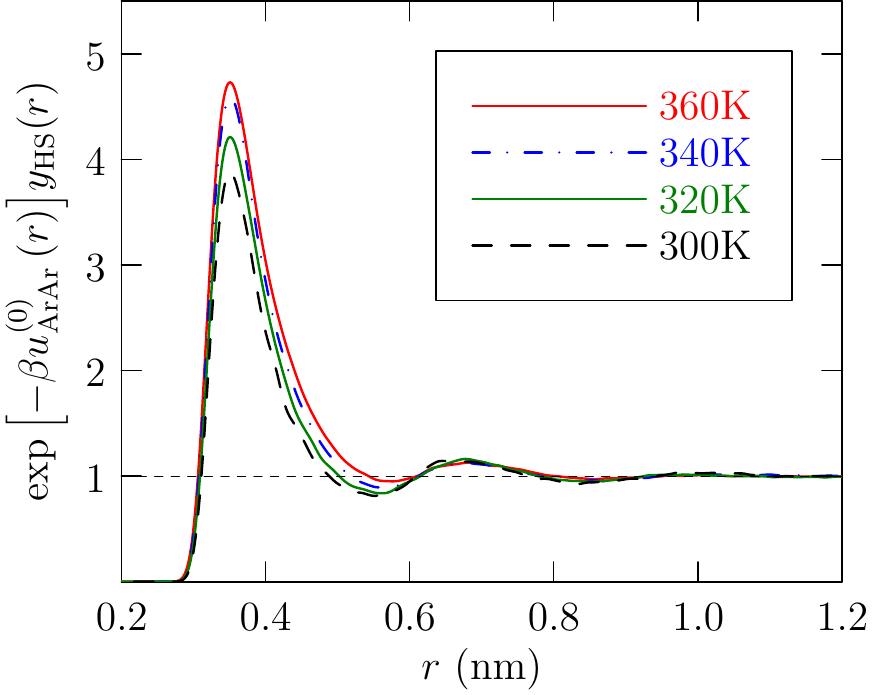}
\caption{Modeled radial distribution functions for WCA repulsive-force
Ar solutes, based on the hard-sphere cavity distribution functions
\cite{chaudhari2013molecular}.}
\label{fg:g0}
\end{figure}

\begin{figure}
\includegraphics[width=3.0in]{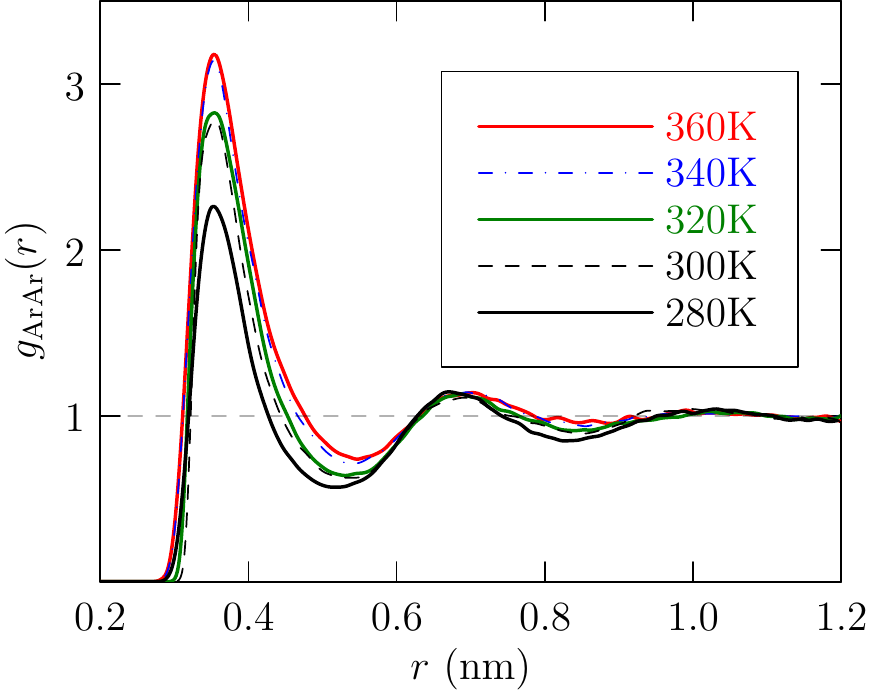}
\caption{Ar-Ar radial distribution functions reconstructed from stratified
(window) calculations. Notice (compare FIG.~\ref{fg:g0}) that contact
hydrophobic interactions are weaker when solute attractive forces are included.
In contrast, solvent-separated correlations are more strongly structured with
inclusion of atomic attractive forces.}
\label{fg:gAA} \end{figure}

Distinctions \cite{MICThesis} deriving from inclusion of solute attractive
forces are exemplified in FIG.~\ref{fg:g0} and FIG.~\ref{fg:gAA}. Inclusion of
solute attractive forces \emph{diminishes} the strength of hydrophobic bonding:
solvent attraction to the solute tends to pull the solute species apart. This
behavior could be expected from sensitive appreciation
\cite{PrattLR:Effsaf,Asthagiri:2008in} of preceding results. The local molecular
field theory (LMF) discussed below is a simple, persuasive theory for these
effects of attractive interactions \cite{jrodg08fd}. Clarifying and testing that
theory is the goal of this work.

Though the substantially the same theory we test below was known and used
\cite{hansen1976theory,PrattLR:Effsaf,PhysRevLett.80.4193} long ago, the insight
underlying recent discussions of LMF theory, \emph{e.g.,} \cite{jrodg08fd}, have
considerably strengthened it. We here give a concise derivation with a clear
analogy to a thermodynamic van der Waals picture, and is therefore unusually
compelling. In the next section we outline the LMF theory. Numerical results,
and conclusions are identified in Sec.~\ref{sec:results}. Methods for the
several computational steps are collected in Sec.~\ref{sec:methods}.

\section{Local molecular field theory} The LMF idea is to study the
inhomogeneous density of a fluid subject to an external field. We focus on the
density structure resulting from the placement of an Ar atom at a specific
location. That distinguished atom exerts an external field on the surrounding
fluid and distorts the density. With $U$ the intermolecular potential energy
function for the system and $\Phi$ the external field exerted by the
distinguished atom, the resulting distorted density is
$\rho_{\alpha\mathrm{M}}\left(\vec{r};U, \Phi\right)$ at position $\vec{r}$ of
$\alpha$ atoms of a molecule of type M. The goal of the LMF theory is to analyze
$\rho_{\alpha\mathrm{M}}\left(\vec{r};U, \Phi\right)$ on the basis of the
characteristics of the interactions $U$ and $\Phi$.

We assume that a reference potential energy, denoted by $U^{(0)}$, has been
identified to help in analyzing
$\rho_{\alpha\mathrm{M}}\left(\vec{r};U, \Phi\right)$. Specifically, our goal is
the match
\begin{eqnarray}
\rho_{\alpha\mathrm{M}}\left(\vec{r};U, \Phi \right) = 
\rho_{\alpha\mathrm{M}} (\vec{r};U^{(0)}, \Phi^{(0)} )~,
\label{eq:match}
\end{eqnarray}
achieved for the reference system with interactions $U^{(0)}$, and an effective
external field $\Phi^{(0)}$. That effective field is the objective
of the analysis below. A successful match Eq.~\eqref{eq:match}
establishes aspects of $U$ that can be treated as molecular mean-fields, thus
offering a molecular mechanism for
$\rho_{\alpha\mathrm{M}}\left(\vec{r};U, \Phi \right)$.

Identification of a reference potential energy function $U^{(0)}$ thus requires
physical insight. One suggestion for the inter-atomic force fields of
current simulation calculations corresponds to Gaussian-truncated electrostatic
interactions associated with the partial charges of simulation
models \cite{jrodg08fd}. In that case, the crucial difference 
\begin{eqnarray}
U-U^{(0)} = 
\sum_{{\alpha\mathrm{M}}, {\gamma\mathrm{M}^\prime}} u_{\alpha\mathrm{M}\gamma\mathrm{M}^\prime}^{(1)}
\left(\vert\vec{r}_{\alpha\mathrm{M}} - \vec{r}_{\gamma\mathrm{M}^\prime}\vert\right)~,
\end{eqnarray}
is atom-pair decomposable. For the case of interest here,
$u_{\alpha\mathrm{M}\gamma\mathrm{M}^\prime}^{(1)}
\left(\vert\vec{r}_{\alpha\mathrm{M}} -
\vec{r}_{\gamma\mathrm{M}^\prime}\vert\right)$ is the WCA-attractive
part of the Lennard-Jones interactions associated with the Ar atoms 
\cite{hansen1976theory}.

In seeking the match Eq.~\eqref{eq:match}, we adopt an atom-based perspective,
and focus on the chemical potential \cite{BPP}, 
\begin{multline}
\mu_{\alpha\mathrm{M}} = 
\beta^{-1}\ln \left\lbrack
\rho_{\alpha\mathrm{M}}\left(\vec{r};U, \Phi\right)\Lambda_{\alpha\mathrm{M}}{}^3 \right\rbrack\\
+ \varphi_{\alpha\mathrm{M}}\left(\vec{r}\right) 
+ \mu_{\alpha\mathrm{M}}^{\mathrm{(ex)}}\left(\vec{r}; \rho, \beta U\right)~,
\label{eq:chemicalpotential}
\end{multline}
of $\alpha$M atoms, which decomposes
\begin{eqnarray}
\Phi = \sum_{\alpha\mathrm{M}} \varphi_{\alpha\mathrm{M}}\left(\vec{r}_{\alpha\mathrm{M}}\right)~.
\end{eqnarray}
Here the temperature is $T = \left(k_{\mathrm{B}}\beta\right)^{-1}$; the thermal 
deBroglie wavelength $\Lambda_{\alpha\mathrm{M}}$ depends only on $T$ and on
fundamental parameters associated with $\alpha$ atoms.  As indicated, the excess
contribution $\mu_{\alpha\mathrm{M}}^{\mathrm{(ex)}}\left(\vec{r}; \rho, \beta
U\right)$ depends functionally on $\left(\rho, \beta U\right)$, not on the
external field.

For some simulation models, the \emph{atom-based} 
 $\mu_{\alpha\mathrm{M}} $ (Eq.~\eqref{eq:chemicalpotential}) may
raise questions regarding the operational status of atom chemical
potentials. But this perspective would be satisfactory for \emph{ab initio}
descriptions of the solution, and is sufficiently basic that we do not further
side-track this discussion.
Similarly for the reference case
\begin{multline}
\mu_{\alpha\mathrm{M}} ^{(0)}= 
\beta^{-1}\ln \left\lbrack
\rho_{\alpha\mathrm{M}}^{(0)}\left(\vec{r}; U^{(0)}, \Phi^{(0)}\right)
\Lambda_{\alpha\mathrm{M}}{}^3 \right\rbrack\\
+ \varphi_{\alpha\mathrm{M}}^{(0)}\left(\vec{r}\right) 
+ \mu_{\alpha\mathrm{M}}^{\mathrm{(ex)}}
\left(\vec{r}; \rho^{(0)}, \beta U^{(0)}\right)~,
\label{eq:chemicalpotentialref}
\end{multline}
with
\begin{eqnarray}
\Phi^{(0)} = \sum_{j} \varphi_{\alpha\mathrm{M}}^{(0)}\left(\vec{r}_j\right)~.
\end{eqnarray}
The forms Eqs.~\eqref{eq:chemicalpotential} and \eqref{eq:chemicalpotentialref}
allows us to express the match Eq.~\eqref{eq:match} as
\begin{multline}
\varphi_{\alpha\mathrm{M}}^{(0)}\left(\vec{r}\right)  = \varphi_{\alpha\mathrm{M}}\left(\vec{r}\right) \\
+ \left\lbrack \mu_{\alpha\mathrm{M}}^{\mathrm{(ex)}}\left(\vec{r}; \rho, \beta U\right)
- \mu_{\alpha\mathrm{M}}^{\mathrm{(ex)}}(\vec{r}; \rho, \beta U^{(0)})\right\rbrack \\
+ \mathrm{constant}~.
\label{eq:chem-mu}
\end{multline}
The bracketed terms in Eq.~\eqref{eq:chem-mu} depend
functionally on the densities, not on the external field.  The \emph{constant}
in Eq.~\eqref{eq:chem-mu} involves the chemical potentials of the 
two systems.

The approximation 
\begin{multline}
 \mu_{\alpha\mathrm{M}}^{\mathrm{(ex)}}\left(\vec{r}; \rho, \beta U\right)
 \approx \mu_{\alpha\mathrm{M}}^{\mathrm{(ex)}}(\vec{r}; \rho, \beta U^{(0)})
 \\ + \sum_{\gamma
\mathrm{M}^\prime}\int \rho_{\gamma\mathrm{M}^\prime}\left(\vec{r}^\prime\right)
 u_{\gamma\mathrm{M}^\prime\alpha\mathrm{M}}^{(1)}\left(\vert\vec{r}^\prime
- \vec{r}\vert\right) \dif \vec{r}^\prime
\label{eq:matchSolution} \end{multline} 
is then  transparently analogous to van der Waals theory
\cite{Widom:1967tz} for the inclusion of attractive interactions, \emph{i.e.,}
$\Delta \mu\approx -2 a \rho,$ with $a$ the van der Waals parameter describing
attractive intermolecular interactions. Transcribing to the case of Ar(aq) at
infinite dilution produces 
\begin{multline}
\varphi_{\mathrm{Ar}}^{(0)}\left(\vec{r}\right) \approx
\varphi_{\mathrm{Ar}}\left(\vec{r}\right) \\ + \int \left\lbrack
\rho_{\mathrm{O}}\left(\vec{r}^\prime\right) - \rho_{\mathrm{O}}\right\rbrack
 u_{\mathrm{OAr}}^{(1)}\left(\vert\vec{r}^\prime - \vec{r}\vert\right) \dif
\vec{r}^\prime ~. \label{eq:matchSolutionAO} 
\end{multline} 
The fields vanish
far from their source, and therefore the constant contribution of
Eq.~\eqref{eq:chem-mu} is accommodated explicitly in
Eq.~\eqref{eq:matchSolutionAO}. This argument matches the results of
Rodgers and Weeks \cite{Rodgers:2008fd} in the several instances they
considered. Derivations that emphasize alternative (electrostatic) interactions
are available elsewhere \cite{Rodgers:2009ht,Rodgers:2011df,Archer:2013bj}.

\begin{figure} 
\includegraphics[width=3.0in]{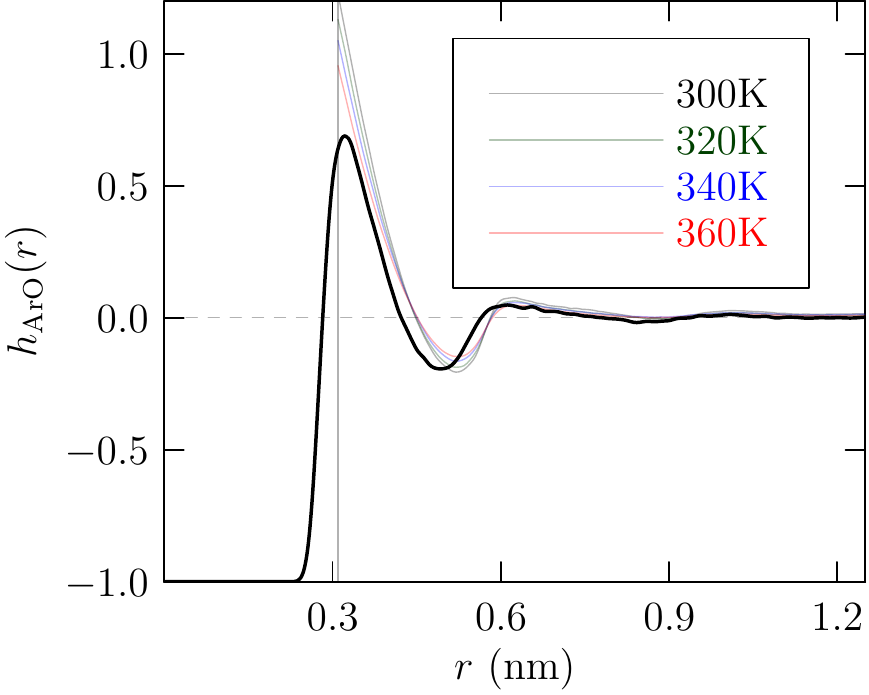}
\caption{ Observed radial correlation of O atoms with an Ar atom, $T$ = 300~K,
$p$ = 1~atm (heavy curve). Correlation functions (fainter, background curves)
for hard-sphere model solutes with distances of closest approach 0.31~nm
(FIG.~\ref{fg:g0}) on the basis of cavity methods
\cite{chaudhari2013molecular,Palma92}, from Chaudhari \cite{MICThesis}. The PC
theory \cite{PRATTLR:Thehe} predictions for the maxima of the hard sphere
correlation functions would be close to 2, larger than these numerical results
\cite{PRATTLR:Thehe,hashb06}. For the soft-sphere case, attractive van der Waals
interactions draw-in near-neighbor O-atoms slightly
\cite{chaudhari2013molecular,AsthagiriD.:NonWth}. Since
attractions draw-in, rather than draw-up, the attractive interactions case is not wetter
than the reference case.} 
\label{fg:hAO} \end{figure}

Though the statistical mechanical theory of Eq.~\eqref{eq:matchSolutionAO} is
simple, the field sought depends on the density, which depends on the field. A
linear statistical mechanical approximation Eq.~\eqref{eq:matchSolution}
produces the non-linear Eq.~\eqref{eq:matchSolutionAO} to solve. The
non-linearity is not an obstacle here because the densities on the right of
Eq.~\eqref{eq:matchSolutionAO} are straightforwardly obtained from routine
simulation (FIG.~\ref{fg:hAO}, see also \cite{Palma92}). Notice
(FIG.~\ref{fg:hAO}) that the effects of attractive ArO interactions on ArO
correlations are modest, as was argued long ago \cite{PrattLR:Effsaf}.

Now consider $\rho_{\mathrm{Ar}} (\vec{r};U^{(0)}, \Phi^{(0)} ),$ the density of
Ar atoms without attractive interactions  $\beta
u_{\mathrm{OAr}}^{(1)}\left(\vec{r} \right)$ but experiencing the effective field
$\beta\varphi_{\mathrm{Ar}}^{(0)}\left(\vec{r}\right).$  We
approximate \cite{Chandler:1970tw}  
\begin{multline} 
 \ln \rho_{\mathrm{Ar}} (\vec{r};U^{(0)}, \Phi^{(0)} ) /\rho_{\mathrm{Ar}}
 \approx -\beta \varphi_\mathrm{Ar} ^{(0)}(\vec{r}) 
 + \ln y_{\mathrm{HS}}(\vec{r})~\\
 = -\beta \left(\varphi_{\mathrm{Ar}}^{(0)}\left(\vec{r} \right) - u_\mathrm{ArAr} ^{(0)}\left(\vec{r} \right) 
\right)
+\ln g_{\mathrm{ArAr}}^{(0)} \left(\vec{r} \right)  ~,
\label{eq:approximate}
\end{multline}
adopting the repulsive-force solute results of FIG.~\ref{fg:g0}. The field
$\varphi_{\mathrm{Ar}}^{(0)}\left(\vec{r} \right)$ incorporates  aspects of
the intermolecular attractions as mean-field effects according to
Eq.~\eqref{eq:matchSolutionAO}. The match Eq.~\eqref{eq:match} 
pairs this with
\begin{equation} 
 \ln \rho_{\mathrm{Ar}} (\vec{r};U, \Phi ) /\rho_{\mathrm{Ar}}
 = \ln g_{\mathrm{ArAr}} \left(\vec{r} \right)~.
 \end{equation}
Combining with Eq.~\eqref{eq:approximate}
\begin{multline} 
\ln g_{\mathrm{ArAr}} \left(\vec{r} \right) = 
\ln g_{\mathrm{ArAr}}^{(0)} \left(\vec{r} \right) 
-\beta \left(\varphi_{\mathrm{Ar}}\left(\vec{r} \right) - u_\mathrm{ArAr} ^{(0)}\left(\vec{r} \right) 
\right) \\
- \int \left\lbrack
\rho_{\mathrm{O}}\left(\vec{r}^\prime\right) - \rho_{\mathrm{O}}\right\rbrack
\beta u_{\mathrm{OAr}}^{(1)}\left(\vert\vec{r}^\prime - \vec{r}\vert\right)
\dif \vec{r}^\prime~.
\end{multline}
Finally noting $\varphi_{\mathrm{Ar}} =  u_\mathrm{ArAr} ^{(0)}+u_\mathrm{ArAr} ^{(1)}$
and rearranging yields
\begin{multline}
-\ln 
\left\lbrack 
\frac{g_{\mathrm{ArAr}}\left(\vec{r}\right)}{g_{\mathrm{ArAr}}^{(0)} \left(\vec{r} \right) }\right\rbrack  
\approx  
\beta u_{\mathrm{ArAr}}^{(1)}\left(\vec{r} \right) \\
 +  \int 
h_{\mathrm{ArO}}\left(\vec{r}^\prime\right)
\rho_{\mathrm{O}}\beta u_{\mathrm{OAr}}^{(1)}\left(\vert\vec{r}^\prime - \vec{r}\vert\right)
\dif \vec{r}^\prime .
\label{eq:LMFapplication}
\end{multline}

\subsection{Comparison to EXP theory \cite{PrattLR:Effsaf}}
As noted above, the approximate theory Eq.~\eqref{eq:LMFapplication} requires
$h_{\mathrm{ArO}}\left(\vec{r}\right),$ and we can conveniently take that from
routine simulation. The corresponding approximate theory for
$h_{\mathrm{ArO}}\left(\vec{r}\right)$ is
\begin{multline}
-\ln 
\left\lbrack 
\frac{g_{\mathrm{ArO}}\left(\vec{r}\right)}{g_{\mathrm{ArO}}^{(0)} \left(\vec{r} \right) }\right\rbrack  
\approx  
\beta u_{\mathrm{OAr}}^{(1)}\left(\vec{r} \right) \\
 +  \int 
h_{\mathrm{OO}}\left(\vec{r}^\prime\right)
\rho_{\mathrm{O}}\beta u_{\mathrm{OAr}}^{(1)}\left(\vert\vec{r}^\prime - \vec{r}\vert\right)
\dif \vec{r}^\prime ,
\label{eq:EXPapplication}
\end{multline}
where $h_{\mathrm{OO}}\left(\vec{r}\right)$ is the observed OO correlation
function for pure water. Acknowledging closure approximations specific to
traditional implementations, this is just the EXP approximation
\cite{hansen1976theory} applied to this correlation problem long ago
\cite{PrattLR:Effsaf,PhysRevLett.80.4193}. This observation serves further to
identify Eq.~\eqref{eq:LMFapplication}  as a relative of the EXP approximation
also. Nevertheless, the distinction between the theory of
Ref.~\cite{PrattLR:Effsaf}, with its specific implementation details, from
Eq.~\eqref{eq:LMFapplication} should be kept in mind. The most prominent distinction is
that Eq.~\eqref{eq:LMFapplication} exploits
$h_{\mathrm{ArO}}\left(\vec{r}\right)$ evaluated self-consistently or, here, the
numerically exact result.

Note further that the Eq.~\eqref{eq:LMFapplication} offers additional
possibilities compared to Eq.~\eqref{eq:EXPapplication} for variety of outcomes
because of possibilities from imbalance of $u_{\mathrm{OAr}}^{(1)}\left(\vec{r}
\right)$ and $ u_{\mathrm{ArAr}}^{(1)}\left(\vec{r} \right) $. 

\subsection{Perspective from quasi-chemical theory
\cite{PrattLR:Effsaf,Asthagiri:2008in,Rogers2012}} Quasi-chemical theory (QCT)
provides insight into the LMF approximation Eq.~\eqref{eq:LMFapplication}. Since
QCT is designed to evaluate interaction contributions to chemical
potentials,\cite{chaudhari2013molecular,AsthagiriD.:NonWth,Sabo:2008dp}
Eq.~\eqref{eq:chem-mu} is the relevant starting point. From the QCT formulation
\cite{Asthagiri:2008in,Rogers2012}, the packing contributions to those two
chemical potentials are identical, and cancel each other. Next to be considered
\cite{Asthagiri:2008in} is the mean hydration energy, denoted by $\left\langle
\varepsilon \vert r, n_\lambda=0\right\rangle$, of the Ar appearing at $r$. That
previous QCT effort \cite{Asthagiri:2008in} observed that the outer-shell QCT
fluctuation contribution was comparatively small. Thus $\left\langle
\varepsilon \vert r, n_\lambda=0\right\rangle$ is the leading factor in
describing the effect of attractive interactions being added
\cite{PrattLR:Effsaf,Asthagiri:2008in}. In the QCT study,\cite{Asthagiri:2008in}
$\left\langle \varepsilon \vert r, n_\lambda=0\right\rangle$ was evaluated from
molecular simulation data. The PC modeling of long-ago \cite{PrattLR:Effsaf}
recognized the importance of $\left\langle \varepsilon \vert r,
n_\lambda=0\right\rangle$, and used a RISM approximation to incorporate the
specific structure of the Ar$_2$ diatom. Returning to the LMF theory, the
right-most of Eq.~\eqref{eq:matchSolution} addresses $\left\langle \varepsilon
\vert r, n_\lambda=0\right\rangle$, but does not calculate it for the detailed
Ar$_2$ geometry. Incomplete evaluation of $\left\langle \varepsilon \vert r,
n_\lambda=0\right\rangle$ is thus the chief neglect of the present LMF theory.

\begin{figure} 
\includegraphics[width=3.0in]{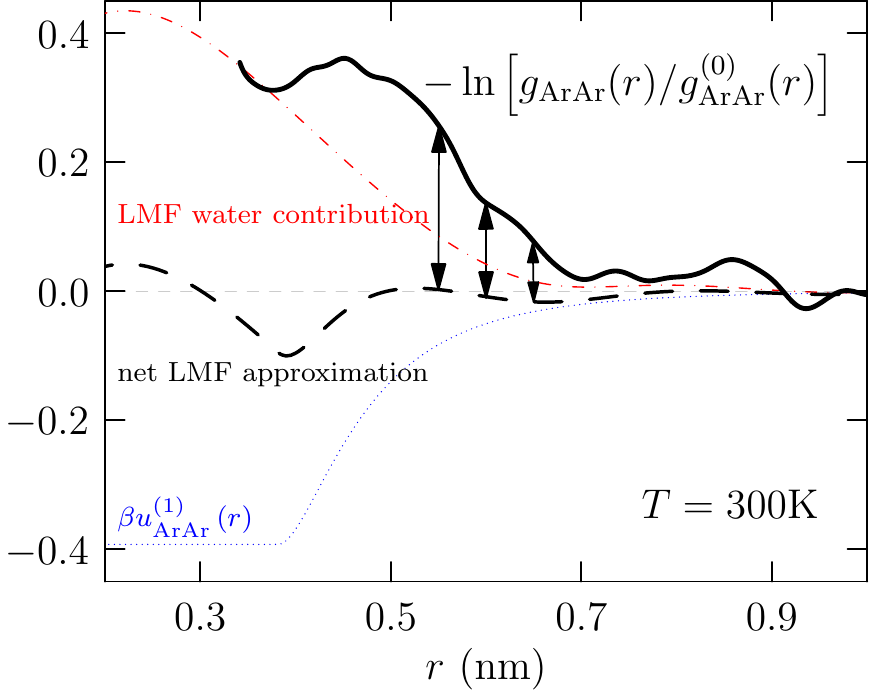} 
\caption{Test of the LMF theory, Eq~\eqref{eq:LMFapplication}. The net result 
for the LMF approximation (black, dashed, right-side of 
Eq.~\eqref{eq:LMFapplication}) is the sum of the direct interaction 
(blue, dotted curve) and the water contribution (red, dotdashed curve,
Eq.~\eqref{eq:bDwAA}). $-\ln\left\lbrack
g_{\mathrm{ArAr}}(r)/g^{(0)}_{\mathrm{ArAr}}(r)\right\rbrack$ differs from
the net LMF approximation both in contact and solvent-separated
configurations.} 
\label{fig:lmf} 
\end{figure}

\begin{figure}
\includegraphics[width=3.0in]{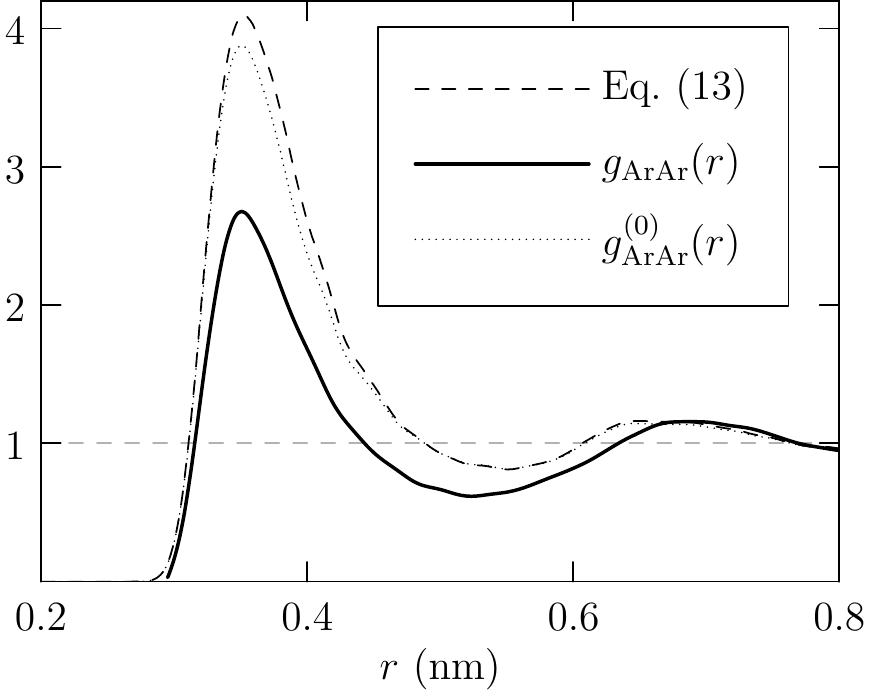}
\caption{Comparison of LMF approximation Eq.~\eqref{eq:LMFapplication} with
$g_{\mathrm{ArAr}}^{(0)} \left(r \right)$ (reference system, fainter, dotted
curve) and $g_{\mathrm{ArAr}}\left(r \right)$. Note the significantly different
behavior of $g_{\mathrm{ArAr}}^{(0)} \left(r \right)$ and
$g_{\mathrm{ArAr}}\left(r \right)$ in the second shell,  not
addressed by this approximation. }
\label{fig:gAALMFCompare} 
\end{figure}

\section{Methods}\label{sec:methods}

\subsection{Simulations} The simulations were carried-out with the GROMACS
package \cite{2008GROMACS}, the SPC/E model of the water molecules
\cite{Berendsen1987}, and the OPLS force field. GROMACS selects SETTLE
\cite{SETTLE} constraint algorithm for rigid SPC/E water molecules. The same
constraint algorithm was used in previous simulations involving
water \cite{chaudhari2013molecular,chaudhari2014hydration}. Standard periodic
boundary conditions were employed, with particle mesh Ewald utilizing a cutoff
of 1~nm and long-range dispersion corrections applied to energy and pressure.
The Parrinello-Rahman barostat controlled the pressure at 1~atm, and the
Nose-Hoover thermostat was used to maintain the temperature. The simulation cell
for the Ar(aq) system consisted of two (2) argon molecules and 1000 water
molecules. Initial configurations were constructed with PACKMOL
\cite{Martinez2009} to construct a system close to the density of interest. The
solute-solute separation spanning 0.33 nm to 1.23 nm was stratified using a
standard windowing approach and the results combined using the weighted histogram
analysis method (WHAM) \cite{Kumar1992}. This involved 19 windows (and
simulations) for window separations $r$ ranging from 0.33~nm to 1.23~nm.

\subsection{Implementation of LMF theory}
With the information of FIG.~\ref{fg:hAO}, the LMF approximation
Eq.~\eqref{eq:LMFapplication} depends only linearly on the attractive
interactions. We evaluated Eq.~\eqref{eq:LMFapplication} standardly,
introducing the spatial Fourier transforms
\begin{eqnarray}
 \hat{u}_{\mathrm{OAr}}^{(1)}\left( k\right) = 
 \int u_{\mathrm{OAr}}^{(1)}\left( r\right) \left(\frac{\sin kr}{kr}\right)  \dif 
 \vec{r}~,
\end{eqnarray}
and 
\begin{eqnarray}
 \hat{h}_{\mathrm{OAr}}\left( k\right) = 
 \int h_{\mathrm{OAr}}\left( r\right) \left(\frac{\sin kr}{kr}\right)  \dif 
 \vec{r}~.
\end{eqnarray}
Then 
\begin{multline}
\int \hat{h}_{\mathrm{ArO}}\left( k\right) \rho_\mathrm{O} \beta\hat{u}_{\mathrm{OAr}}^{(1)}\left( k\right)  
 \left(\frac{\sin kr}{kr}\right)
  \frac{\dif \vec{k}}{\left(2\pi\right)^3 }\\
=   \int 
h_{\mathrm{ArO}}\left(\vec{r}^\prime\right)
\rho_{\mathrm{O}}\beta u_{\mathrm{OAr}}^{(1)}\left(\vert\vec{r}^\prime - \vec{r}\vert\right)
\dif \vec{r}^\prime .
\label{eq:bDwAA}
\end{multline}
The parameters for this application are $\rho_\mathrm{O} =
33.8/\mathrm{nm}^3$, $\varepsilon_{\mathrm{OAr}} = 0.798$~kJ/mol,  
$\sigma_{\mathrm{OAr}} = 0.328$~nm, 
$\varepsilon_{\mathrm{ArAr}} = 0.978$~kJ/mol,  and 
$\sigma_{\mathrm{ArAr}} = 0.340$~nm.

\subsection{Osmotic $B_2$ and Infinite Size Extrapolation}\label{sec:OB2}
The distribution function $g_{\mathrm{ArAr}}\left( r \right) =
h_{\mathrm{ArAr}}\left( r \right) + 1$ provides access to the osmotic second
virial coefficient,
\begin{eqnarray}
  B_2 = -\frac{1}{2}\lim_{\rho_{\mathrm{Ar}}\rightarrow 0}
  \int h_{\mathrm{ArAr}}\left( r \right) \dif^3 r ~.
\end{eqnarray}
We utilize the extrapolation procedure of
Kr\"{u}ger, \emph{et al.} \cite{kruger2012kirkwood,schnell2013apply} 
\begin{eqnarray}
-2B_2 = \lim_{R\rightarrow \infty}
 4\pi \int_0^{2R}  h_{\mathrm{ArAr}}\left( r \right)
 w(r/2R) r^2 \dif r~,
 \label{eq:wKruger}
\end{eqnarray}
with 
\begin{eqnarray}
 w(x) = 1 - \left(\frac{3}{2}\right)x + \left(\frac{1}{2}\right)x^3~.
\end{eqnarray}
Computed values for $1/2R >0$ were least-squares fitted with a polynomial
quadratic order in $1/2R$, then extrapolated to $1/2R =0$. This procedure has
been successfully tested
\cite{chaudhari2014hydration,Ashbaugh:2015cx,WZhang2015} and does not require
further statistical mechanical theory.

\begin{figure}
\includegraphics[width=3.0in]{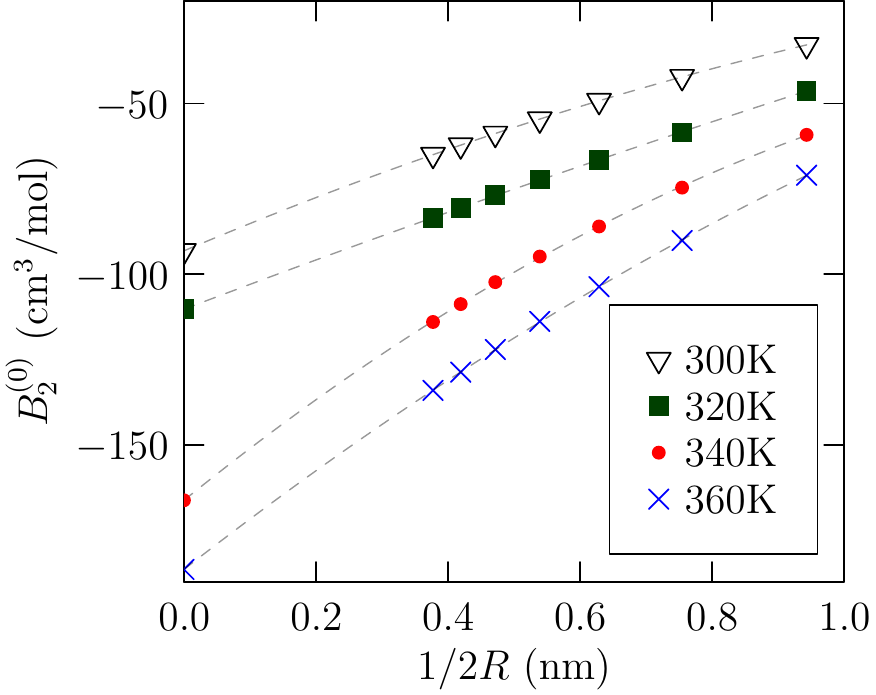}
\caption{Extrapolation to evaluate the osmotic second virial coefficient,
$B_2^{(0)}$, for the WCA repulsive-force Ar solutes (FIG.~\ref{fg:g0}). The
symbol at $1/2R =0$ is the extrapolated value; see Sec~\ref{sec:OB2}.
Hydrophobic interactions gauged by $B_2^{(0)}$ become more attractive with
increasing temperature in this range.}
\label{fig:B2naught} 
\end{figure}

\begin{figure}
\includegraphics[width=3.0in]{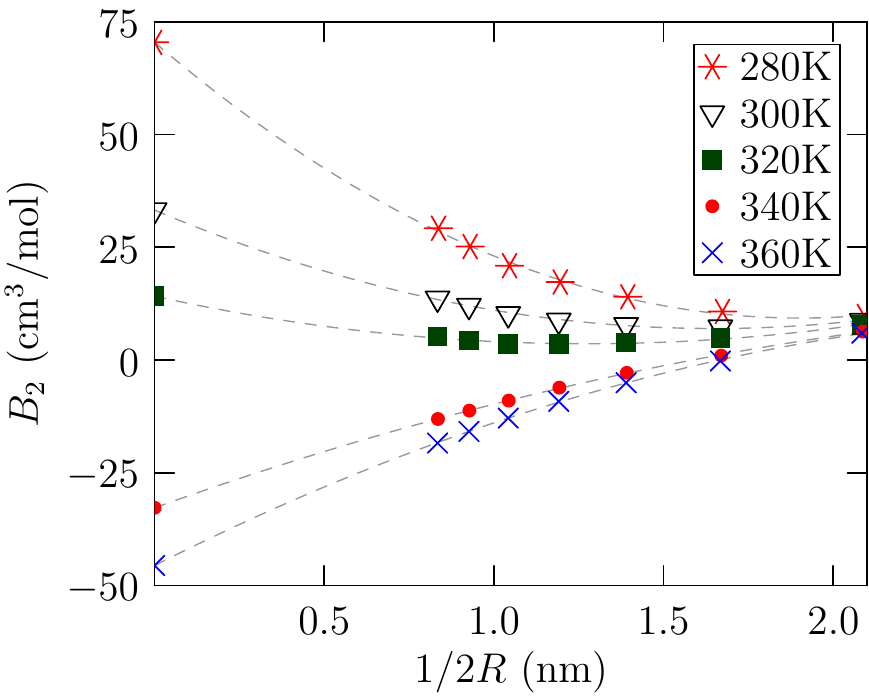}
\caption{Extrapolation (Sec~\ref{sec:OB2}) to evaluate the osmotic second virial
coefficient, $B_2$, for the full attractions case of FIG.~\ref{fg:gAA}.
Comparing with FIG.~\ref{fig:B2naught}, we see that inclusion of solute
attractive-forces makes these $B_2$ more positive (repulsive). Hydrophobic
interactions gauged by $B_2$ become more attractive with increasing temperature
in this range. $B_2$ changes from positive to negative values in $T$
=(320K, 340K). Thus $B_2 \approx 0$ in that interval, qualitatively consistent
with the work of Watanabe, \emph{et al.,} \cite{WATANABEK:MOLSOT}.}
\label{fig:B2} 
\end{figure}

\section{Results and Discussion} \label{sec:results}

Changing purely repulsive atomic interactions to include realistic attractions
\emph{diminishes} the strength of hydrophobic bonds (FIGs.~\ref{fg:g0} and
\ref{fg:gAA}). Within this LMF theory, and also the earliest theories for this
\cite{PrattLR:Effsaf,Asthagiri:2008in}, the hydration environment competes with
direct Ar-Ar attractive interactions (FIG.~\ref{fig:lmf}). The outcome of that
competition is sensitive to the differing strengths of the attractive
interactions. The earlier application \cite{PrattLR:Effsaf} used the EXP
approximation to analyze the available Monte Carlo calculations on atomic LJ
solutes in water \cite{PANGALIC:AMCs}. That theoretical modelling found modest
effects of attractive interactions, and encouraging comparison with the Monte
Carlo results. This application of the LMF theory (Eq.~\ref{eq:LMFapplication})
again predicts modest effects of attractive interactions, but the net comparison
from the simulation results shows big differences. The outcome alternative to
the historical work is due to the fact that the earlier theory used the PC
approximate results for the reference system $g_{\mathrm{ArAr}}^{(0)}\left(
r\right)$, and we now know that approximation is not accurate for this
application \cite{chaudhari2013molecular}, despite being the only theory
available. Here the LMF theory (Eq.~\eqref{eq:LMFapplication}) predicts
modest-sized changes, though in addition it predicts changes opposite in sign to
the observed changes. Note further that $g_{\mathrm{ArAr}}^{(0)}\left( r\right)$
and $g_{\mathrm{ArAr}}\left( r\right)$ differ distinctively in the second
hydration shell, and those differences suggest more basic structural changes
driven by attractive interactions.

The earliest study of these effects \cite{PrattLR:Effsaf} went further to
analyze a Lennard-Jones model of CH$_4$-CH$_4$ (aq) with much larger attractive
interactions. The theory developed for that application was successful for the
case studied \cite{SMITHDE:Freeea}, but the correspondence of that LJ model to
CH$_4$ solutes was not accurate enough \cite{SMITHDE:Freeea} to warrant further
interest.  

That earlier theory featured study of $\left\langle \varepsilon \vert r,
n_\lambda=0\right\rangle$ that has acquired a central role in QCT study of the
present problem \cite{Asthagiri:2008in}. A more accurate evaluation would
involve $n$-body ($n>2$) correlations, perhaps even treated by natural
superposition approximations \cite{Ashbaugh:2005jh}. Detailed treatment of the
Ar$_2$ diatom geometry is the most prominent difference between that QCT
approach and the present LMF theory (Eq.~\eqref{eq:LMFapplication}).
Nevertheless, a full QCT analysis of these differences is clearly warranted and
should be the subject of subsequent study.

These changes due to attractive interatomic interactions are directly reflected
in the values of $B_2$ (FIGs.~\ref{fig:B2naught} and \ref{fig:B2}). Slight
curvature of the extrapolation (FIGs.~\ref{fig:B2naught} and \ref{fig:B2}) is
evident but, in view of the previous testing
\cite{chaudhari2014hydration,Ashbaugh:2015cx,WZhang2015}, not concerning. In all
cases here, $B_2$ becomes more attractive with increasing temperature below $T$
= 360K. This behavior is consistent with accumulated experience and recently
obtained results \cite{chaudhari2013molecular,koga2013osmotic,Ashbaugh:2015cx}.
With attractive interactions in play, $B_2$ can change from positive to negative
values with increasing temperatures. This is consistent with the historical work
of Watanabe, \emph{et al.,} \cite{WATANABEK:MOLSOT} that $B_2 \approx 0$ for
intermediate cases.

Finally, we emphasize that since attractions make substantial contributions,
precise tests of the PC theory \cite{PANGALIC:AMCs,Garde:1996p7972} with results
on cases with realistic attractive interactions should specifically address the
role of attractive interactions that were not included in the PC theory. 

\section{Acknowledgement} We thank J. D. Weeks (University of Maryland) for
helpful discussions. Sandia is a multiprogram laboratory operated by Sandia
Corporation, a Lockheed Martin Company, for the U.S. Department of Energy's
National Nuclear Security Administration under Contract No. DE-AC04-94AL8500.
The financial support of Sandia's LDRD program and the Gulf of Mexico Research
Initiative (Consortium for Ocean Leadership Grant SA 12-05/GoMRI-002) is
gratefully acknowledged.



\providecommand{\latin}[1]{#1}
\providecommand*\mcitethebibliography{\thebibliography}
\csname @ifundefined\endcsname{endmcitethebibliography}
  {\let\endmcitethebibliography\endthebibliography}{}

\end{document}